\documentclass[runningheads]{llncs}

\usepackage{microtype}
\usepackage{iftex}
\iftutex
\else
\usepackage[utf8]{inputenc}
\usepackage[T1]{fontenc}
\fi
\usepackage{newtxtext,newtxmath} 

\usepackage{float} 
\usepackage[normalem]{ulem} 
\usepackage{enumitem} 
\usepackage{needspace} 

\usepackage{amsmath}
\usepackage{stmaryrd}
\usepackage{tikz}
\usetikzlibrary{calc,arrows.meta,bending,nfold}
\usepackage{tikz-cd}
\usepackage[htt]{hyphenat}
\usepackage{cancel} 
\usepackage[dvipsnames]{xcolor}
\usepackage{graphicx} 
\spnewtheorem{exampledef}[example]{Example}{\itshape}{\normalfont}
\spnewtheorem{nonexample}[example]{Non-example}{\itshape}{\normalfont}
\spnewtheorem{notation}{Notation}{\itshape}{\normalfont}
\spnewtheorem{convention}{Convention}{\itshape}{\normalfont}

\newcommand{\phantomsout}[1]{\sout{\phantom{#1}}}
\newcommand*{\etc}{\textit{etc.\@}}
\newcommand*{\opcit}{\textit{op.\@{} cit.\@}}

\newcommand*{\chap}{Ch.\@}
\newcommand*{\sect}{Sec.\@}
\newcommand*{\etal}{~\textit{et al.\@}}

\usepackage{hyperref}  
\iffalse
\newcommand{\linkmarginpar}[1]{%
\ifvmode\relax%
\AddToHookNext{para/begin}{\dolinkmarginpar{#1}}%
\else\dolinkmarginpar{#1}\fi}
\newcommand{\dolinkmarginpar}[1]{\marginpar[\raggedleft\relax#1]{\raggedright\relax#1}}
\else
\usepackage{marginnote} 
\newcommand{\linkmarginpar}[1]{%
\marginnote[\raggedleft\relax#1]{\raggedright\relax#1}}
\fi
\newsavebox{\rocqlogobox}\begin{lrbox}{\rocqlogobox}\includegraphics[height=1.6ex]{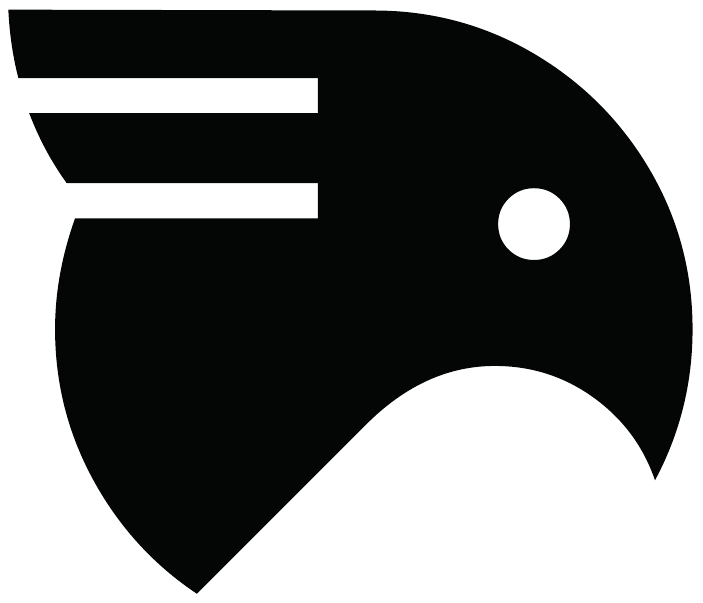}\end{lrbox}
\newcommand*\rocqlogo{\usebox{\rocqlogobox}}

\newcommand{\lemmaref}[3]{\href{https://github.com/UniMath/UniMath/blob/94fae0eda57a778cec704bfa73b4ff525c9bc475/#1\#L#2?name=#3}{\rocqlogo}}
\newcommand{\orgcode}[1]{\textup{\texttt{#1}}}
\newcommand{\acat}[1]{\mathcal{#1}}

\newcommand{\C}{\acat{C}}
\newcommand{\D}{\acat{D}}
\newcommand{\M}{\acat{M}}
\newcommand{\PP}{\acat{P}}
\newcommand{\NN}{\acat{N}}
\newcommand{\OO}{\acat{O}}
\newcommand{\EE}{\acat{E}}

\newcommand{\Cl}[1]{#1_l}
\newcommand{\Ct}[1]{#1_t}
\newcommand{\Ci}[1]{#1_i}
\newcommand{\Cti}[1]{#1_{ti}}
\newcommand{\Cli}[1]{#1_{li}}
\newcommand{\Clt}[1]{#1_{lt}}
\newcommand{\Clti}[1]{#1_{lti}}
\newcommand{\Cn}[1]{#1^\ominus}
\newcommand{\Cp}[1]{#1^\oplus}
\newcommand{\Csn}[2]{#1^{\ominus#2}}
\newcommand{\Csp}[2]{#1^{\oplus#2}}
\newcommand{\Cnl}[1]{\Cn{\Cl{#1}}}
\newcommand{\Cpt}[1]{\Cp{\Ct{#1}}}
\newcommand{\Csnl}[2]{\Csn{\Cl{#1}}{#2}}
\newcommand{\Cspt}[2]{\Csp{\Ct{#1}}{#2}}

\newcommand{\wkcat}[1]{\mathbf{#1}}
\newcommand{\Setcat}{\wkcat{Set}}

\newcommand{\Terminal}{\wkcat{1}}
\newcommand{\Obl}[1]{\OO[#1]}
\newcommand{\Env}[1]{\EE[#1]}
\DeclareMathOperator{\ob}{\mathsf{ob}}

\newcommand{\op}{{\mathsf{op}}}

\newcommand{\id}{\mathsf{id}}
\newcommand{\Id}{\mathsf{Id}}

\newcommand{\dcomp}{\fatsemi}

\newcommand{\morsof}[3]{{#1}\llbracket{#2},{#3}\rrbracket}

\newcommand{\defequiv}{\overset{\mathrm{\textup{def}}}{\equiv}}

\newcommand{\refl}{\mathsf{refl}}
\newcommand{\fib}[1]{\mathsf{efib}_{#1}}

\newcommand{\idtoiso}{\mathsf{id\_to\_iso}} 

\newcommand{\ofan}[2]{\{#2\}^{{\Rightarrow}}_{#1}}
\newcommand{\ifan}[2]{\{#2\}^{{\Leftarrow}}_{#1}}

\newcommand{\force}{\mathop{\mathsf{force}}\nolimits}
\newcommand{\delay}{\mathop{\mathsf{delay}}\nolimits}
\newcommand{\wrap}{\mathop{\mathsf{wrap}}\nolimits}
\newcommand{\unwrap}{\mathop{\mathsf{unwrap}}\nolimits}
\newcommand{\upshift}{\mathord{\Uparrow}}
\newcommand{\downshift}{\mathord{\Downarrow}}

\newcommand{\inneg}{\mathop{\iota^{\ominus}}}
\newcommand{\inpos}{\mathop{\iota^{\oplus}}}
\newcommand{\projneg}{\mathop{\pi^{\ominus}}}
\newcommand{\projpos}{\mathop{\pi^{\oplus}}}
\newcommand{\crossmor}{\mathop{\mathsf{cross}}}
\newcommand{\innegE}{\mathop{\iota^{\ominus}_{\EE}}}
\newcommand{\inposE}{\mathop{\iota^{\oplus}_{\EE}}}
\newcommand{\projnegE}{\mathop{\pi^{\ominus}_{\EE}}}
\newcommand{\projposE}{\mathop{\pi^{\oplus}_{\EE}}}
\newcommand{\crossmorE}{\mathop{\mathsf{cross}_{\EE}}}
\newcommand{\innegO}{\mathop{\iota^{\ominus}_{\OO}}}
\newcommand{\inposO}{\mathop{\iota^{\oplus}_{\OO}}}
\newcommand{\projnegO}{\mathop{\pi^{\ominus}_{\OO}}}
\newcommand{\projposO}{\mathop{\pi^{\oplus}_{\OO}}}
\newcommand{\crossmorO}{\mathop{\mathsf{cross}_{\OO}}}
\newcommand{\pullR}[1]{{#1}^{\triangleright}}
\newcommand{\pullL}[1]{{#1}^{\triangleleft}}

\author{Beatrice Szilvasy\orcidID{0009-0003-2902-6554}}
\date{\today}
\title{Univalent Non\hyp{}Associative Categorical Structures}
\hypersetup{
 pdfauthor={Beatrice Szilvasy\orcidID{0009-0003-2902-6554}},
 pdftitle={Univalent Non\hyp{}Associative Categorical Structures},
 pdfkeywords={},
 pdfsubject={},
 pdfcreator={},
 pdflang={English}}
\begin{document}

\institute{University of Cambridge}

\maketitle
\begin{abstract}
We study category\hyp{}like structures in which composition is not
associative, especially Munch\hyp{}Maccagnoni's duploids, in univalent
foundations. Included is a definition of what a \emph{univalent duploid}
is: one where isomorphisms of a suitable kind render the objects
indistinguishable to the type theory. We show that the existing
construction of a duploid from an adjunction is not univalent as
a ``single\hyp{}sorted'' duploid, and provide a novel construction which
is. This work is formalized in the \textsc{UniMath}
library for the \textsc{Rocq} proof assistant.

\keywords{Univalent Foundations \and Duploids \and Adjunction Models \and Polarization}
\end{abstract}
\section{Introduction}
\label{sec:introduction}
In this paper we study two non\hyp{}associative categorical structures:
\emph{unital magmoids} and \emph{duploids}. A unital magmoid consists of the
same data as a category, but does not require composition to be
associative. A duploid is special kind of unital magmoid in which
objects are ``polarized'' and which is equipped with a ``polarity shift'' structure
\cite{MunchMaccagnoni14}. Polarization in this definition corresponds
to the recovery of associativity for certain composites, as developed
in Munch\hyp{}Maccagnoni's PhD thesis \cite{MunchMaccagnoni13}, but has
many guises.

In the terminology of call\hyp{}by\hyp{}push\hyp{}value \cite{Levy99,Levy17}, \emph{value types}
are positive and \emph{computation types} are negative. Polarity shifts
axiomatize \emph{thunks} -- values which are frozen computations -- and
their dual, \emph{returners} -- computations which return values.

Duploids have been used to study and unify the theory of polarization
in programming languages, (classical and linear) logic, and their
categorical models
\cite{MunchMaccagnoni13,Clairambault17,Mangel25,Mangel26}. As a
pertinent example, where one may construct a Kleisli category from a
monad to model call\hyp{}by\hyp{}value
\cite{Moggi91,MunchMaccagnoni14,Fuhrmann99}, and a co\hyp{}Kleisli
category from a comonad to model call\hyp{}by\hyp{}name
\cite{Mangel25,MunchMaccagnoni13}, one may construct a duploid from
an adjunction which contains both the Kleisli and co\hyp{}Kleisli
categories as subcategories. In fact, every duploid arises this way
\cite{MunchMaccagnoni14}.

\subsubsection{Univalence} In set\hyp{}theory, \emph{equality} (or
\emph{identification}, written \(=\)) is a global equivalence relation
which renders the related objects indistinguishable to the logic. In
univalent foundations, the notion of identification, while still
rendering the objects indistinguishable, is less strict than equality
in set\hyp{}theoretic foundations. In exchange, identification in univalent
foundations is able to coincide with weaker notions of equivalence,
such as isomorphism (of sets, groups, rings, \etc), or adjoint
equivalence of categories.

As an example, it is well\hyp{}known that in a category \(\C\), the correct
notion of equivalence between two objects \(a,b : \C\) is
isomorphism \(a \cong b\), and not necessarily identification \(a =
b\).  In set\hyp{}theoretic foundations, equality is typically overly
restrictive: there may be objects \(a \cong b\) that are not
equal,\footnote{Say, the sets \(\{0,1\}\) and \(\{1,2\}\).} and there may be isomorphisms \(a \cong a\) that are not
the identity.\footnote{Say, the set \(\{\bot,\top\}\) under boolean negation.} In univalent foundations, the type of
identifications \(a =_{\ob\C} b\) may very well correspond to
isomorphisms \(a \cong b\). A \emph{univalent category} is one where they
do \cite{Ahrens15}. As an example, a partial order is a univalent
category, even in set\hyp{}theoretic foundations.

\subsubsection{Contribution} Mathematics in univalent foundations must
be carefully formulated in order to take advantage of the relaxed
notion of identification. For example, one must avoid introducing any
structure which is not preserved under the desired notion of
equivalence. The existing construction of a duploid from an adjunction
fails in this regard.

We carry out the necessary (re)formulation in univalent foundations. For a
start, we define what it means for unital magmoids and duploids to be
\emph{univalent}. We must already be careful, for there is a bifurcation of
definitions of ``duploid'', between ``single\hyp{}sorted'' and ``two\hyp{}sorted''
variants, and they accordingly have different definitions of
univalence.

A unital magmoid is called univalent if certain well\hyp{}behaved
isomorphisms corresponds to identifications. We find that such
isomorphisms in a unital magmoid ought to be not only linear
\cite{MunchMaccagnoni14} and thunkable \cite{Fuhrmann99}, but also
satisfy a third property which we call \emph{intermediate}. In a duploid,
intermediateness is trivialized for isomorphisms that are both linear
and thunkable already.

The original construction of a duploid \(\D\) from an adjunction
\cite{MunchMaccagnoni14,Mangel25} introduces extra structure on
objects that is not invariant under (linear\hyp{}and\hyp{}thunkable)
isomorphism. The issue manifests as non\hyp{}univalence of \(\D\) when
viewed as a unital magmoid or single\hyp{}sorted duploid. Our main result
is the construction of an alternative duploid from an adjunction,
which does satisfy the single\hyp{}sorted univalence condition.

Our work is formalized in the \textsc{Rocq} library \textsc{UniMath}
\cite{UniMath}. This work constitutes a change of 7958 lines of code added across 22 files, as
counted by \textsc{Cloc} \cite{Cloc}. The material of this work is
located in the directory \orgcode{UniMath/CategoryTheory/Nonassociative/}.
The corresponding \textsc{Rocq} code of each definition and proof is
hyperlinked with \rocqlogo{}.

\subsubsection{Related Work} This paper is primarily based on
Munch\hyp{}Maccagnoni's \emph{Models of a Non\hyp{}Associative Composition}
\cite{MunchMaccagnoni14,MunchMaccagnoni13}, which first
introduces duploids. Much of our work is based on formalizing the
definitions and proofs of \opcit{} in \textsc{Rocq}. The parts that we do
not include are the internal language of a duploid, and the full
``structure theorem'' relating the category of duploids to the category
of adjunctions. The rest of Munch\hyp{}Maccagnoni's PhD thesis
\cite{MunchMaccagnoni13} explores non\hyp{}associative composition and
polarization more broadly.

Other work on duploids includes that of Mangel\etal{} \cite{Mangel25}
about linear classical logic, which also includes some general theory
about duploids. We note that our ``correct notion of isomorphism'' in a
unital magmoid (\autoref{rem:um-yoneda-isos}) differs from theirs, but they do
coincide for preduploids (\autoref{lem:preduploid-lt-iso-intermediate}).

Duploids are related to Levy's call\hyp{}by\hyp{}push\hyp{}value and its adjunction
models \cite{Levy99,Levy03b}. Especially pertinent are Levy's
\emph{Contextual Isomorphisms} \cite{Levy17}. Levy defines \emph{contextual
isomorphism} of types \(\theta : A \cong B\) syntactically with
respect to an equivalence relation \(\sim\) on typed terms. Given a
typing judgement with a type hole \(\Gamma[-] \vdash C[-]\), a
contextual isomorphism bijectively maps \(\sim\)-classes of typed
terms \(\Gamma[A] \vdash t : C[A]\) to \(\sim\)-classes of typed terms
\(\Gamma[B] \vdash \theta(t) : C[B]\). Levy shows for
call\hyp{}by\hyp{}push\hyp{}value that when \(\sim\) includes computation rules,
contextual isomorphisms correspond to thunkable (Führmann
\cite{Fuhrmann99}) isomorphisms for value (positive) types, and
linear (Munch\hyp{}Maccagnoni \cite{MunchMaccagnoni14}) isomorphisms for
computation (negative) types. This result is similar to our definition of
univalence for duploids in \autoref{sec:dupl-univalence}, and our
proofs of univalence in \autoref{sec:equalizing}.

For the ``univalent'' part, this paper is based on the study of category
theory in univalent foundations initiated by Ahrens\etal{}
\cite{Ahrens15}. Our work is also based on \opcit{} in the
literal sense that we directly use its \textsc{Rocq} formalization, which
is today part of \textsc{UniMath} \cite{UniMath}. Areas of category
theory already studied in univalent foundations include bicategories
\cite{Ahrens21}, monoidal categories \cite{Wullaert23} and monads
\cite{vanDerWeide25}.

Another work is \emph{The Univalence Principle} by Ahrens\etal{}
\cite{Ahrens22}, deriving algorithmically what it means for a large
class of mathematical structures to be ``univalent''.  Ahrens\etal{}
mention duploids, but use the ``two\hyp{}sorted'' definition rather than the
``single\hyp{}sorted'' one. Still, our definitions of univalence are the same
as derived by their procedure \cite[Example 13.5]{Ahrens22}.
\section{Type Theory}
\label{sec:type-theory}
Our results are written, and to be understood, as informal
mathematical prose. This prose is nevertheless intended to be
interpretable in the formal language of homotopy type theory
\cite{hottbook,Rijke25}. We therefore adopt some vernacular and
notation from the latter, which we note here. Set theorists may, for
intuition, read \emph{``proposition''} as ``subsingleton'', \emph{``equivalence (of
types)''} as ``isomorphism (of sets)'', and \emph{``is a set''} as
tautological.\footnote{Our results are in fact consistent, though less interesting, with uniqueness of identity proofs.\label{org13f6ecc}}

As in \textsc{UniMath} \cite{UniMath}, we work in Martin\hyp{}Löf
type theory \cite{MLTT79} with intensional identity types and no
W\hyp{}types. Unlike \textsc{UniMath}, we do not assume any
universes, let alone a univalent one.\footref{org13f6ecc} We freely assume function
extensionality.

We write products \(\prod_{x:B} E[x]\), sums \(\sum_{x:B} E[x]\),
binary products \(A \times B\), binary sums \(A \amalg B\),
non\hyp{}dependent function types \(A \to B\), and finite types as sets
\(\{\texttt{a},\texttt{b},\dots\}\). Judgemental equality is written with \(\equiv\) and
the identification type with \(=_A\), or simply \(=\).

A type is \emph{contractible} if it has exactly one element, and a
\emph{proposition} if it has at most one element. A type \(A\) is a \emph{set}
if for all \(a,b:A\) the identification type \(a =_A b\) is a
proposition. The \emph{propositional truncation} of a type \(A\) is the
proposition that \(A\) is inhabited. The inhabitant of a type may be
obtained from the propositional truncation so long as the
representative is irrelevant: a principle called \emph{unique choice}.

\begin{convention}
\label{con:proptrunc}
The phrase ``there exists a unique'' is to be read as saying the
subsequent type is contractible. Unlike the \emph{HoTT Book}
\cite{hottbook}, we use propositionally truncated interpretations of
logic: ``exists'' and ``inhabited'' are to be read as propositional
truncation; ``or'' is a binary sum, and under propositional truncation
unless preceded by ``either''.
\end{convention}

\begin{definition}[{\lemmaref{UniMath/Foundations/PartA.v}{1220}{isweq} \cite[\chap{} 4]{hottbook}}]
A function \(f \colon A \to B\) is an \emph{equivalence} (\emph{of types}), written \(f : A \simeq
B\), if for all \(y:B\) there exists a unique \(x:A\) with an
identification \(f(x)=y\). Every equivalence \(f : A \simeq B\) has a
two\hyp{}sided inverse \(f^{-1} \colon B \to A\), and every function with a
two\hyp{}sided inverse may be upgraded to an equivalence.
\end{definition}
\section{Unital Magmoids}
\label{sec:unital-magmoids}
\label{sec:firstbody}
In this section we recall the definition of a univalent category, and
generalize it to unital magmoids. Along the way, we introduce
Munch\hyp{}Maccagnoni's characterization of polarity as the failure of
associativity \cite{MunchMaccagnoni14,MunchMaccagnoni13}, as it is
necessary to define the correct notion of isomorphism in a unital
magmoid.

A unital magmoid consists of the same data as a category
\cite{Ahrens15}, but need not satisfy associativity of
composition. Explicitly:

\begin{definition}[{\lemmaref{UniMath/CategoryTheory/Nonassociative/UnitalMagmoids/Core.v}{126}{unital\_magmoid}}]
\label{def:umagmoid}
A \emph{unital magmoid} or \emph{non\hyp{}associative category} \(\M\) consists of:
\begin{enumerate}
\item A type of \emph{objects}, written \(\ob\M\) or just \(\M\).
\item For all \(a,b : \ob\M\), a type of
\emph{morphisms} from \(a\) to \(b\), also known as a
\emph{hom\hyp{}type}, written \(\morsof\M{a}{b}\) or \(a \to b\).
\item For all \(a : \ob\M\), an \emph{identity} morphism \(\id_a \colon a \to
   a\).
\item A \emph{composition} operator \(\dcomp\) assigning to each pair of
morphisms \(f \colon a \to b\) and \(g \colon b \to c\), a composite
\(f \dcomp g : a \to c\).
\item Composition must be \emph{unital}: \(\id_a \dcomp f = f\) and \(f \dcomp
   \id_b = f\) for all \(f \colon a \to b\).
\item \(\M\) must have \emph{hom\hyp{}sets}: each hom\hyp{}type \(\morsof\C{a}{b}\) is a
set.\footnote{The hom\hyp{}set law is orthogonal to matters of size, and says that
\(f =_{\morsof\M{a}{b}} g\) is a proposition.}
\end{enumerate}
\end{definition}

We shall import the notion of ``isomorphism'' \(a \cong b\) for unital magmoids from
categories directly, even though they are less well\hyp{}behaved. For
example, inverses are not necessarily unique in a unital magmoid. To
obtain a better\hyp{}behaved notion, we must require certain composites to
associate.

\begin{definition}[{\lemmaref{UniMath/CategoryTheory/Nonassociative/UnitalMagmoids/Core.v}{159}{is\_linear}}]
\label{def:polarities}
A path of three morphisms \(a \xrightarrow{f} b \xrightarrow{g} c
\xrightarrow{h} d\) in a unital magmoid is said to \emph{associate} when
\(f \dcomp (g \dcomp h) = (f \dcomp g) \dcomp h\).  We say that:
\begin{itemize}
\item \(f\) is \emph{thunkable} if all triples \((f,g,h)\) associate \cite{MunchMaccagnoni14,Mangel25}.
\item \(g\) is \emph{intermediate} if all triples \((f,g,h)\) associate.
\item \(h\) is \emph{linear} if all triples \((f,g,h)\) associate \cite{MunchMaccagnoni14,Mangel25}.
\item \(b\) is (\emph{semantically}) \emph{negative} if all
morphisms into \(b\) are thunkable \cite{Mangel25}. Equivalently, if all morphisms
out of \(b\) are intermediate.
\item \(c\) is (\emph{semantically}) \emph{positive}
when all morphisms out of \(c\) are linear
\cite{Mangel25}. Equivalently, if all morphisms into \(c\) are
intermediate.
\end{itemize}
These are depicted pictorially in \autoref{fig:polarities}.
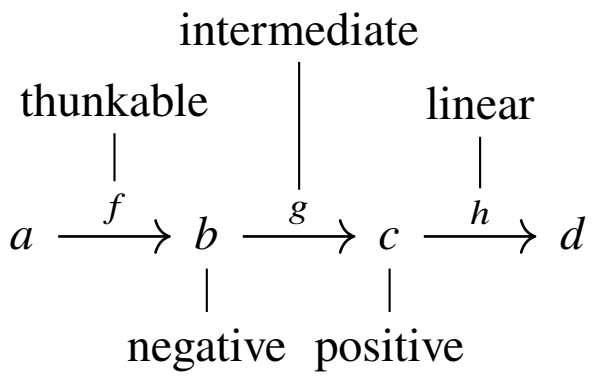
\begin{figure}[H]
\[\begin{tikzpicture}[commutative diagrams/every diagram]
\matrix[matrix of math nodes, name=m, commutative diagrams/every cell] {
a & b & c & d \\};
\path[commutative diagrams/.cd, every arrow, every label]
  (m-1-1) edge node[name=f] {$f$} (m-1-2)
  (m-1-2) edge node[name=g] {$g$} (m-1-3)
  (m-1-3) edge node[name=h] {$h$} (m-1-4);
\draw[font=\footnotesize]
  (f) -- +(0,0.5) node[above] {thunkable}
  (g) -- +(0,1) node[above] {intermediate}
  (h) -- +(0,0.5) node[above] {linear}
  (m-1-2) -- +(0,-0.5) node[below] {negative}
  (m-1-3) -- +(0,-0.5) node[below] {positive};
\end{tikzpicture}\]
\caption{\label{fig:polarities}The triple \((f,g,h)\) associates if any of the labelled properties holds.}
\end{figure}
\end{definition}

\begin{definition}[{\lemmaref{UniMath/CategoryTheory/Nonassociative/UnitalMagmoids/PolarizedSubcategories.v}{86}{linear\_category} \cite{MunchMaccagnoni14}}]
\label{def:subcats}
Let \(\M\) be a unital magmoid. The following are subcategories of
\(\M\) -- that is, unital sub\hyp{}magmoids of \(\M\) which are associative.
\begin{itemize}
\item The \emph{linear subcategory} \(\Cl\M\) of linear morphisms.
\item The \emph{thunkable subcategory} \(\Ct\M\) of thunkable morphisms.
\item The \emph{intermediate subcategory} \(\Ci\M\) of intermediate
morphisms.
\item The (\emph{semantically}) \emph{negative subcategory} \(\Cn\M\) of semantically negative objects.
\item The (\emph{semantically}) \emph{positive subcategory} \(\Cp\M\) of semantically positive objects.
\item Intersections of the above, notably:
\begin{itemize}
\item The \emph{linear\hyp{}and\hyp{}thunkable subcategory} \(\Clt\M\), of linear
and thunkable morphisms.
\item The \emph{linear\hyp{}and\hyp{}thunkable\hyp{}and\hyp{}intermediate subcategory} \(\Clti\M\).
\item The \emph{negative\hyp{}linear subcategory} \(\Cnl\M\).
\item The \emph{positive\hyp{}thunkable subcategory} \(\Cpt\M\).
\end{itemize}
\end{itemize}
\end{definition}

\begin{notation}
\label{not:binop-subcats}
We write the sub/super\hyp{}scripts from \autoref{def:subcats} onto the binary
operators of morphisms (\(\to\)) or isomorphisms (\(\cong\)) to denote
that the morphism or isomorphism (and its inverse) lies in the
corresponding subcategory of the unital magmoid.
\end{notation}
\begin{convention}
Phrasing such as ``linear\hyp{}and\hyp{}thunkable isomorphism'' has the
meaning suggested by \autoref{not:binop-subcats}. That is, a
linear\hyp{}and\hyp{}thunkable isomorphism \(f : a \Clt\cong b\) has that both
\(f \colon a \Clt\to b\) and its inverse \(f^{-1} \colon b \Clt\to a\) are
linear as well as thunkable.
\end{convention}
\subsection{Univalence}
\label{sec:univalence}
\subsubsection*{Internal Univalence}
\label{sec:internal-univalence}
The correct notion of isomorphism in a unital magmoid is a
linear\hyp{}and\hyp{}thunkable\hyp{}and\hyp{}intermediate one.  All three are required for
good behaviour:

\begin{remark}
\label{rem:um-yoneda-isos}
Linearity of \(p : a \Cl\cong b\) corresponds to naturality of
post\hyp{}composition with \(p\), as \((- \dcomp p) : \morsof\M{-}{a}
\simeq \morsof\M{-}{b}\). (Despite \(\morsof\M{-}{a}\) not generally
satisfying functoriality, ``natural transformations'' remain perfectly
well\hyp{}defined on it.) Thunkability is dual. Intermediateness is
required to make \((f \dcomp p) \dcomp (p^{-1} \dcomp g) = f \dcomp
g\) hold, as it is not implied by linearity and thunkability.\footnote{Consider the free unital magmoid \(\wkcat{Lini}\) generated by
morphisms \(a \xrightarrow{f} b \xrightarrow{g} c\) and a
linear\hyp{}and\hyp{}thunkable isomorphism \(p : b \Clt\cong d\). It may be
constructed explicitly with finite types by setting
\(\morsof{\wkcat{Lini}}{a}{c} \defequiv \{\texttt{fg},
\texttt{fpqg}\}\), and all other hom\hyp{}sets to singletons or empty.}
Correctness may also be justified using general techniques:
linear\hyp{}and\hyp{}thunkable\hyp{}and\hyp{}intermediate isomorphisms \(a \Clti\cong b\)
correspond to the indiscernibilities \(a \asymp b\) of
Ahrens\etal{} \cite{Ahrens22}. We therefore define:
\end{remark}

\begin{definition}[{\lemmaref{UniMath/CategoryTheory/Core/Univalence.v}{43}{is\_univalent} \cite{Ahrens15}}]
\label{def:cat-univalent}
A category \(\C\) is \emph{univalent} if the canonical map \((a = b) \to (a
\cong b)\), given by \(\idtoiso\ \refl_a \defequiv \id_a\), is an
equivalence for all \(a,b:\C\).
\end{definition}
\begin{definition}[{\lemmaref{UniMath/CategoryTheory/Nonassociative/UnitalMagmoids/Univalence.v}{121}{is\_um\_univalent}}]
\label{def:um-univalent}
A unital magmoid \(\M\) is \emph{univalent} if its
linear\hyp{}and\hyp{}thunkable\hyp{}and\hyp{}intermediate subcategory is univalent, so
that \((a = b) \simeq (a \Clti\cong b)\) for all \(a,b:\M\).
\end{definition}
\subsubsection*{External Univalence}
\label{sec:external-univalence}
As is typical in category theory, we are not only concerned with
unital magmoids themselves, but also with maps and equivalences
between them. We use the term ``functor'' for maps between unital
magmoids with the same definition as for categories. As usual, a
functor whose morphism part \(F_1 \colon \morsof\NN{a}{b} \to
\morsof\M{F_0\,a}{F_0\,b}\) is an equivalence is called \emph{fully
faithful}. A fully faithful functor is further called an
\emph{isomorphism}, written \(F : \NN \cong \M\), if its object part \(F_0
\colon \ob\NN \to \ob\M\) is also an equivalence of
types. Isomorphisms render the respective categories indistinguishable
to the type theory \cite[Lemma 6.16]{Ahrens15}. For example, we may
derive the following directly:

\begin{theorem}[{\lemmaref{UniMath/CategoryTheory/CategoryRXGraph.v}{214}{is\_univalent\_catiso\_iff}}]
\label{lem:univalence-over-catiso}
If \(\C \cong \D\) for categories, then \(\C\) is univalent if and
only if \(\D\) is.
\qed
\end{theorem}

Isomorphism of categories is not generally as easy to prove as adjoint
equivalence. Between univalent categories, however, the notions
coincide. Univalent categories are even stronger, in fact. It is a
theorem, without the axiom of choice, that for a univalent category
\(\C\) and a category \(\D\), a fully\hyp{}faithful and essentially
surjective functor \(F \colon \C \to \D\) is an adjoint equivalence
\cite{Ahrens15}.  The analogous statement in set\hyp{}theoretic
foundations amounts to the axiom of choice.

We obtain a similar result for unital magmoids in Theorems~
\ref{thm:split-eso-from-eso} and
\ref{thm:is-catiso-from-weak-um-equiv}. In the following, let \(\NN\)
and \(\M\) be unital magmoids and \(F \colon \NN \to \M\) a functor.

\begin{definition}[{\lemmaref{UniMath/CategoryTheory/Nonassociative/UnitalMagmoids/Functors.v}{312}{submm\_essentially\_surjective}}]
\label{def:um-eso}
For \(b : \M\), the type \(\fib{F}(b) \defequiv \sum_{a:\NN} Fa
\Clti\cong b\) is called the (\emph{essential}) \emph{fibre} of \(F\) over
\(b\). The functor \(F\) is \emph{essentially surjective} when all its
fibres are inhabited, and \emph{split essentially surjective} when all its
fibres have a chosen element.
\end{definition}

\begin{definition}[{\lemmaref{UniMath/CategoryTheory/Nonassociative/UnitalMagmoids/Functors.v}{346}{is\_submm\_weak\_equiv}}]
\label{def:um-weak-equiv}
A fully\hyp{}faithful functor is called a \emph{weak equivalence} if it is
essentially surjective, and a \emph{strong equivalence} if it is split
essentially surjective.
\end{definition}

\begin{lemma}[{\lemmaref{UniMath/CategoryTheory/Nonassociative/UnitalMagmoids/Univalence.v}{379}{isaprop\_lti\_fibre}}]
\label{lem:isaprop-lti-fibre}
If \(\NN\) is univalent and \(F\) fully faithful, then all fibres of \(F\) are propositions.
\end{lemma}
\begin{proof}
Elements \(\langle a, Fa \Clti{\overset{e}{\cong}} b \rangle\) and
\(\langle a', Fa' \Clti{\overset{e'}{\cong}} b\rangle\) of
\(\fib{F}(b)\) can be identified by reflecting \(Fa
\Clti{\overset{e}{\cong}} b \Clti{\overset{e'^{-1}}{\cong}} Fa'\) in
\(F\) and using univalence of \(\NN\) to obtain \(p : a = a'\) and
\(F(\idtoiso\ p) = e \dcomp e'\). Induction on \(p\) exposes
\(F(\id_a) = e \dcomp e'^{-1}\) and thus \(e = e'\).
\qed
\end{proof}

\begin{theorem}[{\lemmaref{UniMath/CategoryTheory/Nonassociative/UnitalMagmoids/Univalence.v}{389}{lti\_weak\_equiv\_upgrade}}]
\label{thm:split-eso-from-eso}
If \(\NN\) is univalent and \(F\) a weak equivalence, \(F\) is a
strong equivalence.
\end{theorem}
\begin{proof}
By \autoref{lem:isaprop-lti-fibre} and unique choice.
\qed
\end{proof}

\begin{theorem}[{\lemmaref{UniMath/CategoryTheory/Nonassociative/UnitalMagmoids/Univalence.v}{410}{is\_catiso\_lti\_weak\_equiv}}]
\label{thm:is-catiso-from-weak-um-equiv}
Weak equivalences between univalent \(\NN\) and \(\M\) are
isomorphisms.
\end{theorem}
\begin{proof}
The object part is injective by \autoref{lem:isaprop-lti-fibre} and surjective
by univalence.
\qed
\end{proof}
\section{Duploids}
\label{sec:duploids}
In this section we introduce \emph{duploids}: unital magmoids equipped with
\emph{polarity shifts}. In \autoref{sec:dupl-defs}, we introduce two
variants of duploid: a more recent definition that treats polarities
as a property \cite{Clairambault17,Mangel25}, and one closer to the
original that treats polarities as structure
\cite{MunchMaccagnoni13,MunchMaccagnoni14}. We coin the terms
\emph{single\hyp{}sorted duploid} and \emph{split duploid} for the two. The purpose
of considering both in this paper is that they have different
univalence conditions, and thus different constructions of the \hyperref[sec:from-adjunctions]{Duploid
from an Adjunction}. In \autoref{sec:dupl-univalence} we define what it
means for a duploid to be univalent, and show that it makes polarity
shifts a property rather than structure.
\subsection{Definitions and Properties}
\label{sec:dupl-defs}
\begin{definition}[{\lemmaref{UniMath/CategoryTheory/Nonassociative/Duploids/Core.v}{92}{preduploid}, \lemmaref{UniMath/CategoryTheory/Nonassociative/Duploids/Core.v}{354}{duploid} \cite{MunchMaccagnoni14,Clairambault17,Mangel25}}]
\label{def:duploid}
A (\emph{single\hyp{}sorted}) \emph{preduploid} is a unital magmoid such
that all objects are semantically negative or positive.

A (\emph{single\hyp{}sorted}) \emph{duploid} is a preduploid \(\D\) equipped with \emph{polarity shifts}:
\begin{enumerate}
\item \emph{Negative shifts}. For every object \(a : \D\),
\begin{enumerate}
\item its \emph{upshift}, a semantically negative object written \(\upshift{a} : \Cn\D\),
\item a linear morphism \(\force_a \colon \upshift{}a \Cl\to a\),
\item a linear\hyp{}and\hyp{}thunkable inverse for \(\force_a\) called
\(\delay_a \colon a \Clt\to \upshift{}a\).
\end{enumerate}
\item \emph{Positive shifts}.  Dually, for every object
\(a : \D\),
\begin{enumerate}
\item its \emph{downshift}, a semantically positive object written \(\downshift{}a : \Cp\D\),
\item a thunkable morphism \(\wrap_a \colon a \Ct\to \downshift{a}\),
\item a linear\hyp{}and\hyp{}thunkable inverse for \(\wrap_a\) called
\(\unwrap_a \colon \downshift{}a \Clt\to a\).
\end{enumerate}
\end{enumerate}
\end{definition}

\begin{definition}[{\lemmaref{UniMath/CategoryTheory/Nonassociative/Duploids/Core.v}{191}{split\_preduploid}, \lemmaref{UniMath/CategoryTheory/Nonassociative/Duploids/Core.v}{461}{split\_duploid}}]
A \emph{split preduploid} is a unital magmoid \(\D\) equipped with a
chosen \emph{polarity mapping} \(\varpi \colon \ob\D \to \{\ominus,
\oplus\}\). We say that an object \(a\) is \emph{syntactically
negative} if \(\varpi(a) = \ominus\) and \emph{syntactically positive}
if \(\varpi(a) = \oplus\), and further require that every
\textbf{syntactically} negative/positive object is also \textbf{semantically}
negative/positive.

A \emph{split duploid} is a split preduploid equipped with polarity
shifts, such that the polarity mapping respects the shifts. That
is, all \(\upshift{a}\)/\(\downshift{a}\) must be syntactically
negative/positive.
\end{definition}

\begin{remark}
Every split preduploid is evidently a single\hyp{}sorted
preduploid. Conversely, the law of excluded middle gives a polarity
mapping for any preduploid. For example, by setting \(\varpi(a)
\defequiv \ominus\) if \(a\) is semantically negative, and \(\varpi(a)
\defequiv \oplus\) otherwise. The law of excluded middle does not help
for defining a polarity mapping for duploids, as \(\varpi\) defined
as such will not respect the polarity shifts in general.
\end{remark}

\begin{exampledef}
A category \(\C\) can be viewed as a duploid with polarity shifts
\(\langle\upshift a, \force_a\rangle \defequiv \langle a,
\id_a\rangle\) and \(\langle\downshift a, \wrap_a\rangle \defequiv
\langle a, \id_a\rangle\). If an object \(a : \C\) exists, then \(\C\)
has no definable shift\hyp{}respecting polarity mapping.
\end{exampledef}

\begin{definition}[{\lemmaref{UniMath/CategoryTheory/Nonassociative/Duploids/PolarizedSubcategories.v}{43}{syn\_positive\_category}}]
\label{def:chosen-subcats}
Let \(\D\) be a split preduploid with polarity mapping \(\varpi \colon
\ob\D \to \{\ominus, \oplus\}\). Extending \autoref{def:subcats}, there are the
following subcategories of \(\D\).
\begin{itemize}
\item The \emph{syntactically negative subcategory} \(\Csn\D\varpi\) of syntactically
negative objects.
\item The \emph{syntactically positive subcategory} \(\Csp\D\varpi\) of syntactically
positive objects.
\item Intersections with those of \autoref{def:subcats}, notably:
\begin{itemize}
\item The \emph{syntactically negative\hyp{}linear subcategory}
\(\Csnl\D\varpi\).
\item The \emph{syntactically positive\hyp{}thunkable subcategory}
\(\Cspt\D\varpi\).
\end{itemize}
\end{itemize}
\end{definition}

\begin{remark}
\label{rem:twosorted}
\autoref{def:duploid} makes the polarities in a preduploid merely a property on
unital magmoids, rather than structure as Munch\hyp{}Maccagnoni
\cite{MunchMaccagnoni14} first defined it. The original
\emph{two\hyp{}sorted} duploids of \opcit{} are
precisely the split duploids \(\D\) where \(\langle \upshift{a},
\force_a \rangle \equiv \langle a, \id_a \rangle\) for syntactically
negative \(a : \D\), and \(\langle \downshift{b}, \wrap_b \rangle
\equiv \langle b, \id_b \rangle\) for syntactically positive \(b :
\D\). Any split duploid's shifts can be strictified to this form.
\end{remark}

We note the following characterizations of polarity in a duploid.

\begin{theorem}[{\lemmaref{UniMath/CategoryTheory/Nonassociative/Duploids/Core.v}{596}{is\_thunkable\_iff\_delay\_wrap} \cite[Proposition 13]{MunchMaccagnoni14}}]
\label{lem:thunkable-and-linear-chars}
In a duploid, a morphism \(f \colon a \to b\) is thunkable if and only
if the triple \(a \xrightarrow{f} b \Clt{\xrightarrow{\delay_b}}
\upshift{b} \Cti{\xrightarrow{\wrap_{\upshift{b}}}}
\upshift{\downshift{b}}\) associates. Dually, \(f \colon a \to b\) is
linear if and only if the triple
\(\upshift{\downshift{a}} \Cli{\xrightarrow{\force_{\downshift{a}}}}
\downshift{a} \Clt{\xrightarrow{\unwrap_{a}}} a \xrightarrow{f} b\)
associates.
\qed
\end{theorem}

\begin{theorem}[{\lemmaref{UniMath/CategoryTheory/Nonassociative/UnitalMagmoids/PolarizedSubcategories.v}{750}{is\_negative\_of\_lt\_iso}}]
\label{lem:ltiso-polarities}
Let \(p : a \Clt\cong b\) be a linear\hyp{}and\hyp{}thunkable isomorphism; then
\(a\) is positive if and only if \(b\) is positive, and \(a\) is
negative if and only if \(b\) is.
\qed
\end{theorem}

\begin{corollary}[{\lemmaref{UniMath/CategoryTheory/Nonassociative/Duploids/Core.v}{745}{lt\_iso\_upshift\_of\_negative}}]
\label{lem:positive-and-negative-chars}
An object in a duploid \(a : \D\) is negative if and only if
\(\force_a \colon \upshift{a} \Cl\to a\) is thunkable. Dually, \(a : \D\)
is positive if and only if \(\wrap_a \colon a \Ct\to \downshift{a}\) is linear.
\qed
\end{corollary}
\subsection{Univalence}
\label{sec:dupl-univalence}
Since single\hyp{}sorted preduploids are just unital magmoids with an extra
property, their univalence condition should be the same. However, we
can simplify the notion of isomorphism used. We also find that
univalence of preduploids has a useful characterization in terms of
the polarized subcategories.

\begin{definition}[{\lemmaref{UniMath/CategoryTheory/Nonassociative/Duploids/Univalence.v}{57}{is\_dupl\_univalent}}]
\label{def:duploid-univalence}
A duploid or preduploid \(\D\) is \emph{univalent} (\emph{as a single\hyp{}sorted
(pre)duploid}) if its linear\hyp{}and\hyp{}thunkable subcategory \(\Clt\D\) is
univalent.
\end{definition}

\begin{theorem}[{\lemmaref{UniMath/CategoryTheory/Nonassociative/UnitalMagmoids/PolarizedSubcategories.v}{773}{lt\_iso\_is\_intermediate\_from\_polarized}}]
\label{lem:lt-iso-i-polarized}
If an object \(a\) in a unital magmoid is negative or positive, then
all linear\hyp{}and\hyp{}thunkable isomorphisms \(p : a \Clt\cong b\) are
intermediate.
\end{theorem}
\begin{proof}
By \autoref{lem:ltiso-polarities}, \(a\) and \(b\) share a polarity.
Intermediateness of \(p\) is immediate.
\qed
\end{proof}

\begin{corollary}[{\lemmaref{UniMath/CategoryTheory/Nonassociative/Duploids/Univalence.v}{67}{is\_um\_univalent\_preduploid\_iff}}]
\label{lem:preduploid-lt-iso-intermediate}
Every linear\hyp{}and\hyp{}thunkable isomorphism in a preduploid is
intermediate. Thus, \autoref{def:duploid-univalence} coincides with univalence as
a unital magmoid.
\qed
\end{corollary}

We now obtain a characterization of univalence for preduploids in
terms of its polarized subcategories. We shall use a specialization of
Rijke's \emph{fundamental theorem of identity types} \cite{Rijke25} for
the proof.

\begin{definition}[{\lemmaref{UniMath/IdentitySystems/RXGraph.v}{99}{edges\_from}}]
\label{def:fans}
For an object \(a : \C\) in a category, the type \(\ofan\C{a}
\defequiv \sum_{b:\C} a \cong b\) is called the \emph{outward fan} of
\(a\), and \(\ifan\C{a} \defequiv \sum_{b:\C} b \cong a\) is called
the \emph{inward fan} of \(a\).
\end{definition}

\begin{theorem}[{\lemmaref{UniMath/IdentitySystems/RXGraph.v}{136}{is\_univalent\_from\_isaprop\_edges\_from} \cite[\chap{} 11]{Rijke25}}]
\label{lem:fun-thm-of-ids}
A category is univalent if and only if all its outward (dually,
inward) fans are propositions.
\qed
\end{theorem}

\begin{theorem}[{\lemmaref{UniMath/CategoryTheory/Nonassociative/Duploids/Univalence.v}{692}{is\_dupl\_univalent\_from\_polarized}}]
\label{thm:duploid-univ-of-subcats}
A preduploid \(\D\) is univalent if and only if \(\Cnl\D\) and
\(\Cpt\D\) are univalent.
\end{theorem}
\begin{proof}
The ``only if'' direction follows from fully\hyp{}faithfulness of the
inclusion functors \(\Cnl\D \hookrightarrow \Clt\D \hookleftarrow
\Cpt\D\).  For the ``if'' direction, we apply \autoref{lem:fun-thm-of-ids} and
show that every fan \(\ofan{\Clt\D}{a}\) is a proposition. Using the preduploid
property, if \(a : \D\) is negative, the fan \(\ofan{\Clt\D}{a}\) is
equivalent to the fan \(\ofan{\Cnl\D}{a}\) by \autoref{lem:ltiso-polarities}.
The latter is a proposition by assumption. Dually, \(\ofan{\Clt\D}{a} \simeq
\ofan{\Cpt\D}{a}\) if \(a\) is positive.
\qed
\end{proof}

For split (pre)duploids, \autoref{def:duploid-univalence} is incorrect in
general, due to the fact that objects may be both negative and
positive semantically, but not syntactically. Instead, we use a
syntactic variant of \autoref{thm:duploid-univ-of-subcats} as the definition of
univalence for split duploids.

\begin{theorem}[{\lemmaref{UniMath/CategoryTheory/Nonassociative/Duploids/Univalence.v}{707}{neg\_is\_dupl\_univalent\_if\_split}}]
\label{lem:not-univalent-split-duploid}
If a split duploid has an object which is both semantically negative
and semantically positive, then its underlying single\hyp{}sorted duploid
is not univalent.
\end{theorem}
\begin{proof}
Univalence applied to \(\upshift{a} \Clt\cong \downshift{a}\) would
imply \(\ominus = \varpi(\upshift{a}) = \varpi(\downshift{a}) =
\oplus\).
\qed
\end{proof}

\begin{definition}[{\lemmaref{UniMath/CategoryTheory/Nonassociative/Duploids/Univalence.v}{91}{is\_split\_dupl\_univalent}}]
\label{def:split-duploid-univalence}
A split duploid or preduploid \(\D\) is \emph{univalent} (\emph{as a split
duploid}) if both its syntactically negative\hyp{}linear subcategory
\(\Csnl\D\varpi\) and and its syntactically positive\hyp{}thunkable
subcategory \(\Cspt\D\varpi\) are univalent.
\end{definition}
\begin{remark}
\autoref{def:split-duploid-univalence} coincides with the indiscernibilities
derived for two\hyp{}sorted duploids by Ahrens\etal{} \cite[Example 13.5]{Ahrens22}.
\end{remark}

Objects in category theory are often unique up to canonical
isomorphism. In these cases, univalent categories make them literally
unique; that is, propositions.  The same holds for unital magmoids,
and polarity shifts are one such example.

\begin{lemma}[{\lemmaref{UniMath/CategoryTheory/Nonassociative/UnitalMagmoids/Univalence.v}{137}{isaprop\_linear\_isos\_from\_negative}}]
\label{thm:um-universal-properties}
For a univalent unital magmoid \(\M\) and object \(a : \M\), we have
that the types \(\sum_{b:\Cn\M}(b \Cl\cong a)\) and \(\sum_{b:\Cp\M}(a
\Ct\cong b)\) are propositions.
\end{lemma}
\begin{proof}
Let \(\langle b, b \Cl{\overset{e}\cong} a \rangle\) and \(\langle b', b' \Cl{\overset{e'}\cong} a \rangle\) be of the former type. Then by
\autoref{lem:fun-thm-of-ids} we have \(\langle b, e \dcomp e'^{-1} \rangle
=_{\ifan{\Clti\M}{b'}} \langle b', \id_{b'} \rangle\). It follows that
\(b = b'\) and \(e = e'\).
\qed
\end{proof}

\begin{theorem}[{\lemmaref{UniMath/CategoryTheory/Nonassociative/Duploids/Univalence.v}{390}{isaprop\_has\_polarity\_shifts}}]
\label{thm:isaprop-shifts}
For a univalent preduploid, polarity shifts are unique if they exist.
\end{theorem}
\begin{proof}
The data \(\langle \upshift{a}, \force_a \rangle\) and \(\langle
\downshift{a}, \wrap_a \rangle\) are of the types in
\autoref{thm:um-universal-properties}.
\qed
\end{proof}

\begin{remark}
\label{rem:um-representability}
\autoref{thm:um-universal-properties} suggests a definition of representability
for unital magmoids. In unital magmoids, natural isomorphisms
\(\morsof\M{-}{a} \cong \morsof\M{-}{b}\) are linear isomorphisms \(a
\Cl\cong b\) (by \autoref{rem:um-yoneda-isos}); so let us call a non\hyp{}functorial map
\(F : \M^\op \to \Setcat\) \emph{representable} if there is a negative
object \(b : \Cn\D\) satisfying \(\morsof\M{-}{b} \cong F\), and
dually with positive objects. In that sense, the polarity shift
structure of duploids amounts to saying that all \(\morsof\D{-}{b}\)
and \(\morsof\D{a}{-}\) are representable, by \(\langle \upshift{b},
\force_b \rangle\) and \(\langle \downshift{a}, \wrap_a \rangle\)
respectively.
\end{remark}
\section{Duploids from Adjunctions}
\label{sec:from-adjunctions}
\label{sec:lastbody}
Duploids are related to adjunctions: every adjunction gives rise to a
duploid, and every duploid is equivalent to one that arises from an
adjunction. The duploid arising from an adjunction was first
constructed by Munch\hyp{}Maccagnoni \cite{MunchMaccagnoni14} as a
two\hyp{}sorted duploid, which (by \autoref{lem:not-univalent-split-duploid}) fails
to be univalent as a single\hyp{}sorted duploid in general. We give this
construction the name \emph{oblique duploid}, and seek to give an
equivalent single\hyp{}sorted construction that can be univalent. We shall
explain the new construction, the \emph{envelope duploid}, in
subsections~\ref{sec:construction}--\ref{sec:structure-thm}. Finally,
in \autoref{sec:equalizing} we shall prove its univalence.
\subsection{Construction}
\label{sec:construction}
\subsubsection*{Morphisms}
\label{sec:morphisms}
Recall that an adjunction \(L \dashv R : \NN \to \PP\) may be
viewed as an isomorphism:
\begin{equation*}
\varphi_{p,n} : \morsof\NN{Lp}{n} \simeq \morsof\PP{p}{Rn}
\qquad \text{natural in \(p : \PP^\op\) and \(n : \NN\).} \label{eq:nathomweq}
\end{equation*}
Like Levy \cite{Levy03b}, we call the morphisms on either side of
\(\varphi_{p,n}\) the \emph{oblique morphisms}, written
\(\morsof\OO{p}{n}\). Rather than making an arbitrary choice of the
side, however, we use the following symmetric definition:

\begin{definition}[{\lemmaref{UniMath/CategoryTheory/Nonassociative/Duploids/FromAdjunction/ObliqueMor.v}{46}{oblique\_mor}}]
\label{def:obliquemor}
An \emph{oblique morphism} \(f : \morsof\OO{p}{n}\) consists of a \emph{negative
mate} \(f^\flat : \morsof{\NN}{Lp}{n}\) and a \emph{positive mate}
\(f^\sharp : \morsof{\PP}{p}{Rn}\), satisfying
\(\varphi_{p,n}(f^\flat) = f^\sharp\). The family \(\morsof\OO{p}{n}\)
extends to a functor \(\OO : \PP^\op \times \NN \to \Setcat\) by
pre/post\hyp{}composition in \(\PP\) and \(\NN\).
\end{definition}
\subsubsection*{Objects}
\label{sec:objects}
Our morphisms shall be given by oblique morphisms. But what should the
objects be? A first approach may be to use objects from either \(\NN\)
or \(\PP\).

\begin{definition}[{\lemmaref{UniMath/CategoryTheory/Nonassociative/Duploids/FromAdjunction/Oblique.v}{58}{oblique\_ob}}]
\label{def:oblique-ob}
An \emph{oblique object} \(a : \Obl{L \dashv R}\) is either \(\innegO{n}\)
for \(n : \NN\) or \(\inposO{p}\) for \(p : \PP\). For an oblique
object \(a\), define the \emph{negative} and \emph{positive projections},
\(\projnegO{a} : \NN\) and \(\projposO{a} : \PP\) in the obvious way,
as
\begin{align*}
  \projnegO(\innegO{n}) &\defequiv n & \projposO(\innegO{n}) &\defequiv Rn \\
  \projnegO(\inposO{p}) &\defequiv Lp & \projposO(\inposO{p}) &\defequiv p\mathpunct.
\end{align*}
The morphisms between oblique objects \(a\) and \(b\) are defined as
\(\morsof\OO{\projposO{a}}{\projnegO{b}}\). This is the start of the
definition of the \emph{oblique duploid}.
\end{definition}

In light of \autoref{lem:not-univalent-split-duploid}, this approach has an
unfortunate deficiency: it creates extra structure by making a
polarity mapping \(\varpi(\iota^{\varepsilon}_{\OO}(x)) \defequiv
\varepsilon : \{\ominus, \oplus\}\) definable.\footnote{The ``extra structure'' is apparent if the type \(\NN \amalg
\PP\) is expressed as \(\sum_{\varepsilon : \{\ominus,
\oplus\}}E(\varepsilon)\), setting \(E(\ominus) \defequiv \ob\NN\) and
\(E(\oplus) \defequiv \ob\PP\). From this perspective, the polarity is
an extra data field!} For an
alternative definition without a definable polarity mapping, we must
use different objects.

Let us consider what the objects need. Reasoning backwards, to define
morphisms \(a \to b\) in terms of \(\morsof\OO{p}{n}\), we shall need
to associate some \(p : \PP\) and some \(n : \NN\) to all objects
\(a,b\). Next, in order to uniformly define identities, it stands to
reason that the \(p : \PP\) and \(n : \NN\) associated to an object
\(a\) ought to be related by a morphism \(\morsof\OO{p}{n}\) that will
become \(\id_a\). We call the data \(p : \PP\), \(n : \NN\) and \(f :
\morsof\OO{p}{n}\) an \emph{envelope pre\hyp{}object}; \(f\) will later need an
extra property to behave like an identity.

\begin{definition}[{\lemmaref{UniMath/CategoryTheory/Nonassociative/Duploids/FromAdjunction/Envelope.v}{56}{envelope\_preob}}]
\label{def:envelope-preob}
An \emph{envelope pre\hyp{}object} \(a\) in \(L \dashv R : \NN \to \PP\) is a triple of an object
\(\projnegE{a} : \NN\), an object \(\projposE{a} : \PP\), and a
morphism \(\crossmorE{a} :
\morsof\OO{\projpos{a}}{\projneg{a}}\). We may write an envelope pre\hyp{}object \(a\) as simply the diagram of
its oblique morphism:
\begin{align*}
\projposE{a} \xrightarrow{\crossmorE{a}} \projnegE{a}\mathpunct.
\end{align*}
The morphisms between two envelope pre\hyp{}objects \(a\) and \(b\), are
given by \(\morsof\OO{\projposE{a}}{\projnegE{b}}\).
\end{definition}
\begin{notation}
The operator \(\crossmor\) should be read with a higher precedence
than \(\flat\) and \(\sharp\). That is, all three of \((\crossmor
a)^\flat\), \(\crossmor(a)^\flat\), and \(\crossmor a^\flat\) mean the
same thing.
\end{notation}

\begin{definition}[{\lemmaref{UniMath/CategoryTheory/Nonassociative/Duploids/FromAdjunction/Envelope.v}{337}{envelope\_preob\_of\_positive}}]
\label{def:oblique-cross-mor}
To tie envelope pre\hyp{}objects back to the oblique objects, we may
associate to each oblique object \(a : \Obl{L \dashv R}\) an envelope
pre\hyp{}object \(\widehat{a}\): \[\left(\projposE{\widehat{a}}
\xrightarrow{\crossmorE{\widehat{a}}} \projnegE{\widehat{a}}\right)
\defequiv \left(\projposO{a} \xrightarrow{\crossmorO{a}}
\projnegO{a}\right)\mathpunct,\] where \(\crossmorO{a} :
\morsof\OO{\projposO{a}}{\projnegO{a}}\) is defined as
\begin{align*}
\crossmorO(\innegO{n})^\flat &\defequiv LRn \xrightarrow{\epsilon_n} n
& \crossmorO(\inposO{p})^\flat &\defequiv Rn \xrightarrow{\id_{Rn}} Rn \\
\crossmorO(\innegO{n})^\sharp &\defequiv Lp \xrightarrow{\id_{Lp}} Lp
& \crossmorO(\inposO{p})^\sharp &\defequiv p \xrightarrow{\eta_p} RLp \mathpunct.
\end{align*}
\end{definition}

Turning the definition of \(\widehat{a}\) around, note that oblique
objects are precisely the envelope pre\hyp{}objects \(b\) for which either
\(\crossmorE{b}^\sharp\) or \(\crossmorE{b}^\flat\) is the
identity. Thus, to lose the polarity mapping structure, we could
propositionally truncate away the aforementioned condition. Of course,
internal to the type theory, we cannot compare an arbitrary morphism
to the identity: \(\crossmorE{b}^\flat
=_{\morsof\NN{L(\projposE{b})}{\textbf{??}}}  \id_{L(\projposE{b})}\)
is ill\hyp{}typed! Instead, we ask for \(\crossmorE{b}^\sharp\) or
\(\crossmorE{b}^\flat\) to be an \emph{isomorphism} -- an inverse for
either is called a \emph{polarity choice}. This is the inspiration behind
the envelope duploid.

\begin{definition}[{\lemmaref{UniMath/CategoryTheory/Nonassociative/Duploids/FromAdjunction/Envelope.v}{81}{envelope\_polarization\_choice}}]
A \emph{polarity choice} for an envelope pre\hyp{}object \(b\) is an
inverse for either mate of \(\crossmorE b\). That is, either an inverse
\((\crossmorE b^\sharp)^{-1} \colon R(\projnegE{b}) \to \projposE{b}\)
for \(\crossmorE b^\sharp\) --- a \emph{negative polarity choice}; or
one \((\crossmorE b^\flat)^{-1} \colon \projnegE{b} \to
L(\projposE{b})\) for \(\crossmorE b^\flat\) --- a \emph{positive
polarity choice}.
\end{definition}

\begin{definition}[{\lemmaref{UniMath/CategoryTheory/Nonassociative/Duploids/FromAdjunction/Envelope.v}{329}{envelope\_ob}}]
\label{def:envelope-ob}
An \emph{envelope object} \(a : \Env{L \dashv R}\) is an envelope
pre\hyp{}object such that there exists a polarity choice for
\(a\). That is, \(\crossmorE{a}^\sharp\) or \(\crossmorE{a}^\flat\) is
an isomorphism.
\end{definition}
\begin{remark}
\label{rem:envelope-ob}
Beware \autoref{con:proptrunc}: the polarity choice for each envelope object
is under propositional truncation. Importantly, it does not partake in
identifications of envelope objects, and may only be obtained
constructively if the choice is irrelevant.
\end{remark}

\begin{exampledef}
Every oblique object is clearly an envelope object: each \(\innegE{n}
\defequiv \widehat{\innegO{n}}\) has a negative polarity choice,
and each \(\innegE{p} \defequiv \widehat{\inposO{p}}\) a positive one.
\end{exampledef}
\begin{notation}
We shall freely omit the subscripts \(\EE\) from \(\inposE\),
\(\innegE\), \(\projposE\), \(\projnegE\), and \(\crossmorE\), and
likewise for the \(\OO\) variants. The conflation between \(\EE\) and
\(\OO\) variants is intentional, to encourage viewing oblique objects
as envelope objects.
\end{notation}
\subsubsection*{Composition}
\label{sec:composition}
With the objects well\hyp{}defined, we now seek to define the composite \(f
\dcomp g : \morsof\OO{\projpos a}{\projneg c}\) of oblique morphisms
\(f : \morsof\OO{\projpos a}{\projneg b}\) and \(g :
\morsof\OO{\projpos b}{\projneg c}\) between envelope objects. There
is no immediate way of composing \(f\) and \(g\) in either \(\NN\) or
\(\PP\). Instead, as for the original oblique duploid, the composite
will depend on the polarity of the intermediate object
\cite{MunchMaccagnoni14}. What does the polarity give us?

\begin{definition}
\label{def:envelope-mor-pull}
If an envelope pre\hyp{}object \(b\) has a negative polarity choice,
there exists for all \(g : \morsof\OO{p}{\projneg b}\) a unique
\(\pullL{g} \defequiv g^\sharp \dcomp (\crossmor b^\sharp)^{-1} :
\morsof\PP{p}{\projpos b}\) with \(\pullL{g} \dcomp \crossmor
b^\sharp = g^\sharp\):
\begin{align*}
\begin{tikzcd}[ampersand replacement=\&]
	p \& \\
	{\projpos b} \& {R(\projneg b)}
	\arrow["{\exists!\pullL{g}}"', dashed, from=1-1, to=2-1]
	\arrow["{g^\sharp}", from=1-1, to=2-2]
	\arrow["{\crossmor b^\sharp}"', from=2-1, to=2-2]
\end{tikzcd} \qquad \morsof\OO{p}{\projneg b} &\simeq \morsof\PP{p}{\projpos b} &&\text{if \(b\) has a negative choice.}
\end{align*}
Dually, if an envelope pre\hyp{}object \(b\) has a positive polarity
choice, there exists for all \(f : \morsof\OO{\projpos{b}}{n}\) a
unique \(\pullR{f} \defequiv (\crossmor b^\flat)^{-1} \dcomp
f^\flat : \morsof\NN{\projneg{b}}{n}\) with \(\crossmor b^\flat
\dcomp \pullR{f} = f^\flat\):
\begin{align*}
\begin{tikzcd}[ampersand replacement=\&]
	{L(\projpos b)} \& {\projneg b} \\
	\& n
	\arrow["{\crossmor b^\flat}", from=1-1, to=1-2]
	\arrow["{f^\flat}"', from=1-1, to=2-2]
	\arrow["{\exists!\pullR{f}}", dashed, from=1-2, to=2-2]
\end{tikzcd} \qquad \morsof\OO{\projpos b}{n} &\simeq
\morsof\NN{\projneg b}{n} &&\text{if \(b\) has positive choice.}
\end{align*}
\end{definition}

We may now resolve the composite \(f \dcomp g : \morsof\OO{\projpos
a}{\projneg c}\).

\begin{definition}[{\lemmaref{UniMath/CategoryTheory/Nonassociative/Duploids/FromAdjunction/Envelope.v}{189}{envelope\_compose'}}]
\label{def:envelope-compose-choice}
Let \(a\), \(b\), and
\(c\) be envelope pre\hyp{}objects in \(L \dashv R\), and let \(f :
\morsof\OO{\projpos{a}}{\projneg{b}}\) and \(g :
\morsof\OO{\projpos{b}}{\projneg{c}}\) be oblique morphisms between
them. Given a negative polarity choice for \(b\), define \(f \dcomp g
\defequiv \morsof\OO{\pullL{f}}{\projneg c}(g)\), whose mates are depicted:
\begin{equation}
\begin{tikzcd}[column sep=large, ampersand replacement=\&]
  {L(\projpos a)} \& \\
  {L(\projpos b)} \& {\projneg b} \\
  \& {\projneg c}
  \arrow["{f^\flat}", from=1-1, to=2-2]
  \arrow["{\crossmor b^\flat}"{description}, from=2-1, to=2-2]
  \arrow["{g^\flat}"', from=2-1, to=3-2]
  \arrow["{L(\pullL{f})}"', dashed, from=1-1, to=2-1]
\end{tikzcd}
\qquad
\begin{tikzcd}[column sep=large, ampersand replacement=\&]
  {\projpos a} \& \\
  {\projpos b} \& {R(\projneg b)} \\
  \& {R(\projneg c)}
  \arrow["{f^\sharp}", from=1-1, to=2-2]
  \arrow["{\crossmor b^\sharp}"{description}, from=2-1, to=2-2]
  \arrow["{g^\sharp}"', from=2-1, to=3-2]
  \arrow["{\pullL{f}}"', dashed, from=1-1, to=2-1]
\end{tikzcd} \label{dia:envelope-comp-negative}
\end{equation}
Dually, given a positive
polarity choice for \(b\), define \(f \dcomp g \defequiv
\morsof\OO{\projpos a}{\pullR{g}}(f)\):
\begin{equation}
\begin{tikzcd}[column sep=large, ampersand replacement=\&]
{L(\projpos a)} \& \\
{L(\projpos b)} \& {\projneg b} \\
\& {\projneg c}
\arrow["{f^\flat}", from=1-1, to=2-2]
\arrow["{\crossmor b^\flat}"{description}, from=2-1, to=2-2]
\arrow["{g^\flat}"', from=2-1, to=3-2]
\arrow["{\pullR{g}}", dashed, from=2-2, to=3-2]
\end{tikzcd} \qquad \begin{tikzcd}[column sep=large, ampersand replacement=\&]
{\projpos a} \& \\
{\projpos b} \& {R(\projneg b)} \\
\& {R(\projneg c)}
\arrow["{f^\sharp}", from=1-1, to=2-2]
\arrow["{\crossmor b^\sharp}"{description}, from=2-1, to=2-2]
\arrow["{g^\sharp}"', from=2-1, to=3-2]
\arrow["{R(\pullR{g})}", dashed, from=2-2, to=3-2]
\end{tikzcd} \label{dia:envelope-comp-positive}
\end{equation}
\end{definition}

\begin{remark}[{\lemmaref{UniMath/CategoryTheory/Nonassociative/Duploids/FromAdjunction/Envelope.v}{229}{envelope\_compose\_irrel}}]
\label{thm:envelope-compose-irrel}
If \(b\) in \(a \xrightarrow{f} b \xrightarrow{g} c\) has both a
negative and a positive polarity choice, then pasting of
Diagrams~\ref{dia:envelope-comp-negative} and
\ref{dia:envelope-comp-positive} reveals that both ways of composing
\(f \dcomp g \colon a \to c\) in \autoref{def:envelope-compose-choice} are
equal. This is enough to justify (with respect to \autoref{rem:envelope-ob})
using the polarity choice that exists for each envelope object.
\end{remark}

We now have the underlying unital magmoid of the envelope
duploid:

\begin{definition}[{\lemmaref{UniMath/CategoryTheory/Nonassociative/Duploids/FromAdjunction/Envelope.v}{398}{envelope\_unital\_magmoid}}]
\label{def:envelope-umagmoid}
The \emph{envelope duploid}-to\hyp{}be \(\Env{L \dashv R}\) over \(L \dashv
R : \NN \to \PP\) has:
\begin{enumerate}
\item The objects \(\ob\Env{L \dashv R}\) are envelope objects (\autoref{def:envelope-ob}).
\item Morphisms \(\morsof{\Env{L \dashv R}}{a}{b}\) are the oblique
morphisms \(\morsof\OO{\projpos{a}}{\projneg{b}}\).
\item Composition is as in \autoref{def:envelope-compose-choice}, using the
intermediate object's polarity choice.
\item For an object \(a : \Env{L \dashv R}\), its identity is given by
\(\id_{a} \defequiv \crossmor{a}\).
\item The universal property of \autoref{def:envelope-mor-pull} is precisely
that composition is unital.
\end{enumerate}
\end{definition}

We may also define the oblique duploid by reference to the envelope duploid.

\begin{definition}[{\lemmaref{UniMath/CategoryTheory/Nonassociative/Duploids/FromAdjunction/Oblique.v}{90}{oblique\_unital\_magmoid}}]
\label{def:oblique-umagmoid}
The \emph{oblique duploid}-to\hyp{}be \(\Env{L \dashv R}\) over \(L \dashv
R\) has the following:
\begin{enumerate}
\item The objects \(\ob\Obl{L \dashv R} \defequiv \NN \amalg \PP\) are
oblique objects (\autoref{def:oblique-ob}).
\item Morphisms, composition, and identities are those of the envelope
duploid, where each oblique object is viewed as an envelope object
(\autoref{def:oblique-cross-mor}).
\end{enumerate}
\end{definition}
\subsubsection*{Polarities and Shifts}
\label{sec:polarities-and-shifts}
We now seek to show that the envelope structure is a duploid. First,
we must show it to be a preduploid: that each object is semantically
negative or positive.

\begin{lemma}[{\lemmaref{UniMath/CategoryTheory/Nonassociative/Duploids/FromAdjunction/Envelope.v}{435}{is\_negative\_of\_envelope\_chosen\_negative}}]
\label{lem:envelope-syn-valid}
Objects in the envelope duploid\hyp{}to\hyp{}be \(\Env{L \dashv R}\) with a
negative polarity choice are negative, and those with a positive
polarity choice are positive.
\qed
\end{lemma}

\begin{corollary}[{\lemmaref{UniMath/CategoryTheory/Nonassociative/Duploids/FromAdjunction/Envelope.v}{470}{envelope\_preduploid}}]
Every object of the envelope duploid\hyp{}to\hyp{}be is negative or
positive based on its associated polarity choice. Thus, \(\Env{L
\dashv R}\) is a single\hyp{}sorted preduploid.
\qed
\end{corollary}

\begin{corollary}[{\lemmaref{UniMath/CategoryTheory/Nonassociative/Duploids/FromAdjunction/Oblique.v}{137}{oblique\_polarity\_mapping}, \lemmaref{UniMath/CategoryTheory/Nonassociative/Duploids/FromAdjunction/Oblique.v}{262}{oblique\_preduploid}}]
The mapping \(\varpi(\iota^{\varepsilon}_{\OO}(x)) \defequiv
\varepsilon : \{\ominus, \oplus\}\) is a valid polarity mapping for
the oblique duploid\hyp{}to\hyp{}be. Thus, \(\Obl{L \dashv R}\) is a
split preduploid.
\qed
\end{corollary}

We now seek to define polarity shifts. Since each envelope object
comes associated with an object from each of the negative and positive
categories, it is straightforward to find for each a semantically
negative or positive object in the envelope duploid.

\begin{definition}[{\lemmaref{UniMath/CategoryTheory/Nonassociative/Duploids/FromAdjunction/Envelope.v}{477}{envelope\_upshift}}]
\label{def:envelope-shifts}
Let \(a : \Env{L \dashv R}\) be an envelope object. Define its
polarity shifts as:
\begin{align*}
\upshift{a} &\defequiv \inneg(\projneg a) : \Env{L \dashv R}
 & \downshift{a} &\defequiv \inpos(\projpos a) : \Env{L \dashv R} \\
\force_a &\defequiv \crossmor(\upshift{a}) : \morsof{\Env{L \dashv R}}{\upshift{a}}{a}
 & \wrap_a &\defequiv \crossmor(\downshift{a}) : \morsof{\Env{L \dashv R}}{a}{\downshift{a}} \\
\delay_a &\defequiv \crossmor{a} : \morsof{\Env{L \dashv R}}{a}{\upshift{a}}
 & \unwrap_a &\defequiv \crossmor{a} : \morsof{\Env{L \dashv R}}{\downshift{a}}{a}\mathpunct.
\end{align*}
We leave it to the reader to verify the axioms from \autoref{def:duploid}.
\end{definition}

This completes the construction of the envelope and oblique duploids:

\begin{theorem}[{\lemmaref{UniMath/CategoryTheory/Nonassociative/Duploids/FromAdjunction/Envelope.v}{672}{envelope\_duploid}}]
The \emph{envelope duploid} \(\Env{L \dashv R}\) of
\autoref{def:envelope-umagmoid} is a single\hyp{}sorted duploid, with polarity shifts
given by \autoref{def:envelope-shifts}.
\qed
\end{theorem}

\begin{theorem}[{\lemmaref{UniMath/CategoryTheory/Nonassociative/Duploids/FromAdjunction/Oblique.v}{269}{oblique\_duploid}}]
The \emph{oblique duploid} \(\Obl{L \dashv R}\) of
\autoref{def:oblique-umagmoid} is a split duploid, with polarity shifts given by
\autoref{def:envelope-shifts}.
\qed
\end{theorem}
\subsection{Polarities in the Envelope Duploid}
\label{sec:polarities-in-the-envelope-duploid}
Before we prove univalence of the construction from \autoref{sec:construction}, we
note some important theorems, starting with characterizations of
linearity, thunkability, and polarities following from
\autoref{lem:thunkable-and-linear-chars}. In the following, let \(L
\dashv_{(\eta,\epsilon)} R : \NN \to \PP\) be an adjunction.

\begin{lemma}[{\lemmaref{UniMath/CategoryTheory/Nonassociative/Duploids/FromAdjunction/Envelope.v}{768}{is\_thunkable\_from\_downshift\_iff\_envelope\_unit\_postcompose} \cite[Proposition 14]{MunchMaccagnoni14}}]
\label{lem:envelope-thunkable-char}
A morphism \(f \colon a \to \inposE{p}\) in \(\Env{L \dashv R}\) is
thunkable if and only if \(f^\sharp \dcomp RL\eta_p = f^\sharp \dcomp
\eta_{RLp}\) holds. Dually, a morphism \(g \colon \innegE{n} \to b\)
in \(\Env{L \dashv R}\) is linear if and only if \(LR\epsilon_n \dcomp
g^\flat = \epsilon_{LRn} \dcomp g^\flat\) holds.
\qed
\end{lemma}

\begin{theorem}[{\lemmaref{UniMath/CategoryTheory/Nonassociative/Duploids/FromAdjunction/Envelope.v}{836}{is\_negative\_envelope\_downshift\_iff\_pre\_fixed\_point}}]
\label{lem:envelope-polarities-char}
For all \(p : \PP\), the object \(\inpos{p}\) is negative in \(\Env{L
\dashv R}\) if and only if the equation \(RL\eta_p = \eta_{RLp}\) holds.
Dually, for \(n : \NN\), the object \(\inneg{n}\) is positive in
\(\Env{L \dashv R}\) if and only if the equation \(LR\epsilon_n =
\epsilon_{LRn}\) holds.
\end{theorem}
\begin{proof}
By, \autoref{lem:positive-and-negative-chars}, it is sufficient to check if
\(\force\) is linear or \(\wrap\) is thunkable to see that an object
is negative or positive. The statement then follows from
\autoref{lem:envelope-thunkable-char}.
\qed
\end{proof}

\begin{remark}
\label{rem:envelope-syn-polarities-char}
We may compare the characterization of polarities in
\autoref{lem:envelope-polarities-char} to the statement of semantic polarities
from polarity choices in \autoref{lem:envelope-syn-valid}. Specialized to
\(\inpos{p}\) and \(\inneg{n}\), \autoref{lem:envelope-syn-valid} states the
following:

\begin{quote}
For all \(p : \PP\), the object \(\inpos{p}\) is negative in \(\Env{L
\dashv R}\) if \(\eta_p\) is an isomorphism. Dually, for \(n : \NN\),
the object \(\inneg{n}\) is positive in \(\Env{L \dashv R}\) if
\(\epsilon_n\) is an isomorphism.
\end{quote}

The converse of \autoref{lem:envelope-syn-valid} does not hold in general.\footnote{Consider the unique functor \(\mathbf! : \Setcat \to \Terminal\) to the terminal
category, and the terminal\hyp{}object functor \(1 : \Terminal \to \Setcat\)
which is its right adjoint.  All objects in \(\Env{\mathbf! \dashv
1}\) are (trivially) both negative and positive, but the unit \(!_a
\colon a \to 1\) is not an isomorphism at e.g. \(a \equiv \{\bot, \top\}\).}
However, it does hold for a useful class of adjunctions (see
\autoref{lem:chosen-negative-iff-negative} later).
\end{remark}

We may now provide concrete examples of split duploids which are not
univalent as single\hyp{}sorted duploids. The following examples may be checked
using \autoref{rem:envelope-syn-polarities-char}.

\begin{exampledef}
\label{ex:oblique-nonunivalent-1}
Let \(\C\) be a category with an object \(a : \C\). Then \(\inneg{a}\)
and \(\inpos{a}\) are both semantically negative and positive in the
oblique duploid over the identity adjunction \(\Obl{\Id_{\C} \dashv
\Id_{\C}}\). Thus the latter is not univalent as a single\hyp{}sorted
duploid by \autoref{lem:not-univalent-split-duploid}.
\end{exampledef}

\begin{exampledef}
Consider a fixed inhabited set \(s : \Setcat\), and the exponential adjunction
\((-\times s) \dashv (-)^{\displaystyle s} : \Setcat \to \Setcat\). The
empty set \(\inneg\{\}\) or \(\inpos\{\}\) is both semantically negative and positive
in \(\Obl{(-\times s) \dashv (-)^{\displaystyle s}}\).
\end{exampledef}
\subsection{The Structure Theorem}
\label{sec:structure-thm}
Every two\hyp{}sorted duploid is isomorphic to the oblique duploid over
some adjunction \cite[Proposition II.35]{MunchMaccagnoni13}. We have
the same for the envelope duploid, but with equivalence of unital
magmoids instead of isomorphism. We also show that the oblique and
envelope duploids are equivalent as unital magmoids.

\begin{theorem}[{\lemmaref{UniMath/CategoryTheory/Nonassociative/Duploids/FromAdjunction/Oblique.v}{281}{oblique\_to\_envelope}}]
\label{lem:weak-equivalence}
The duploid \(\Obl{L \dashv R}\) is weakly equivalent to \(\Env{L \dashv R}\).
\end{theorem}
\begin{proof}
The mapping \(a \mapsto \widehat{a}\) on objects \(\Obl{L \dashv R}
\to \Env{L \dashv R}\) extends trivially to a functor. To see that it
is a weak equivalence, observe that every \(b : \Env{L \dashv R}\) is
linearly\hyp{}and\hyp{}thunkably isomorphic to \(\upshift{b} \equiv
\widehat{\inpos(\projpos{b})}\) or to \(\downshift{b} \equiv
\widehat{\inpos(\projneg{b})}\) depending on the polarity choice of
\(b\).
\qed
\end{proof}

\begin{remark}[{\lemmaref{UniMath/CategoryTheory/Nonassociative/Duploids/PolarizedSubcategories.v}{600}{adjunction\_neg\_l\_to\_pos\_t} \cite[Proposition 21]{MunchMaccagnoni14}}]
\label{rem:dupl-adjunction}
The polarity shifts of a duploid \(\D\) extend to morphisms, as
\(\upshift(a \xrightarrow{f} b) \defequiv \force_a \dcomp f \dcomp
\delay_b\) and \(\downshift(a \xrightarrow{f} b) \defequiv \unwrap_a
\dcomp f \dcomp \wrap_b\) (in general, non\hyp{}functorially). These are
functorial when restricted to \(\downshift : \Cnl\D \to \Cpt\D\) and
\(\upshift : \Cpt\D \to \Cnl\D\), which take part in an adjunction
\(\upshift \dashv_{(\eta, \epsilon)} \downshift : \Cnl\D \to \Cpt\D\)
with unit \(\eta_p \defequiv \delay_p \dcomp \wrap_{\upshift p}\) and
counit \(\epsilon_n \defequiv \force_{\downshift n} \dcomp
\unwrap_n\). We then have:
\end{remark}

\begin{theorem}[{\lemmaref{UniMath/CategoryTheory/Nonassociative/Duploids/FromAdjunction/StructureTheorem.v}{359}{weak\_equiv\_duploid\_into\_envelope}}]
\label{thm:envelope-structure}
The duploid \(\D\) is weakly equivalent to \(\Env{\upshift \dashv \downshift : \Cnl\D \to \Cpt\D}\).
\end{theorem}

\begin{proof}
The functor \(F : \D \to \Env{\upshift \dashv \downshift}\) is
determined by its action on morphisms
\begin{align*}
F_1\!\left(a \xrightarrow{f} b\right) &: \morsof\OO{\downshift{a}}{\upshift{b}} \\
F_1\!\left(a \xrightarrow{f} b\right)^\flat
&\defequiv \upshift\!\left(\downshift{a} \xrightarrow{\unwrap_a} a \xrightarrow{f} b\right) \\
F_1\!\left(a \xrightarrow{f} b\right)^\sharp
&\defequiv \downshift\!\left(a \xrightarrow{f} b \xrightarrow{\delay_a} \upshift{b}\right)\mathpunct.
\end{align*}
On objects, we then have \(F_0(a) \defequiv (\downshift{a}
\xrightarrow{F_1(\id_a)} \upshift{a})\). The fibre over each \(b :
\Env{\upshift \dashv \downshift}\) is inhabited with object part
\(\projneg b\) or \(\projpos b\) depending on the polarity of
\(b\). If \(b\) is negative, we must find an isomorphism
\(F(\projneg b) \Clt\cong b\). We have \(\force_b : \upshift b \Clt\cong
b\) and, since \(F\) can be shown to preserve polarities,
\(\delay_{F(\projneg b)} : F(\projneg b) \Clt\cong \upshift F(\projneg
b)\).  We may then compute \(\upshift{F(\projneg b)} \equiv \inneg(\upshift
(\projneg b))\) and \(\inneg(\projneg b) \equiv \upshift b\) to obtain
\(\inneg(\force_{\projneg b}) : \inneg(\upshift (\projneg b))
\Clt\cong \inneg(\projneg b)\). Ultimately, the isomorphism part is:
\begin{equation*}
F(\projneg b)
\overset{\delay}{\Clt\cong} \upshift F(\projneg b)
\equiv \inneg(\upshift (\projneg b))
\overset{\inneg(\force)}{\Clt\cong} \inneg(\projneg b) \equiv \upshift b
\overset{\force}{\Clt\cong} b\mathpunct. \eqno\qed
\end{equation*}
\end{proof}

\begin{remark}[{\lemmaref{UniMath/CategoryTheory/Nonassociative/Duploids/FromAdjunction/StructureTheorem.v}{351}{is\_strong\_equiv\_duploid\_into\_envelope\_from\_LEM}, \lemmaref{UniMath/CategoryTheory/Nonassociative/Duploids/FromAdjunction/Oblique.v}{307}{split\_eso\_oblique\_to\_envelope\_from\_LEM}}]
The equivalences of Theorems~\ref{lem:weak-equivalence} and
\ref{thm:envelope-structure} are strong when assuming the law of
excluded middle. This is because we can ``decide'' whether a negative
polarity choice for \(b : \Env{L \dashv R}\) exists, and take an
arbitrary bias towards \(\projneg{b}\) if it does.
\end{remark}
\subsection{Univalence and the Equalizing Requirement}
\label{sec:equalizing}
For the rest of the section, let \(L \dashv_{(\eta, \epsilon)} R : \NN \to \PP\) be an
adjunction. If \(\NN\) and \(\PP\) are univalent, when are the oblique
and envelope duploids of \(L \dashv R\) univalent? We must somehow
show that \(\Csnl{\Obl{L \dashv R}}\varpi/\Cspt{\Obl{L \dashv
R}}\varpi\) and \(\Cnl{\Env{L \dashv R}}/\Cpt{\Env{L \dashv R}}\) are
univalent. We shall do so using \autoref{lem:univalence-over-catiso}, by
exhibiting isomorphisms of categories with \(\NN/\PP\). To do so, we
will need an \emph{equalizing requirement} \cite{MunchMaccagnoni14,Moggi91}.

\begin{lemma}[{\lemmaref{UniMath/CategoryTheory/Nonassociative/Duploids/FromAdjunction/Oblique.v}{446}{negative\_to\_oblique\_duploid}, \lemmaref{UniMath/CategoryTheory/Nonassociative/Duploids/FromAdjunction/Envelope.v}{901}{negative\_to\_envelope\_duploid}}]
(Recall that \(\Csnl{\Obl{L \dashv R}}\varpi\) is the \textbf{syntactic}
negative\hyp{}linear subcategory, and \(\Cnl{\Env{L \dashv R}}\) the
\textbf{semantic} negative\hyp{}linear subcategory.) The inclusions \(\innegO\),
\(\inposO\), \(\innegE\), and \(\inposE\) extend to functors:
\begin{align*}
\innegO &\colon \NN \to \Csnl{\Obl{L \dashv R}}\varpi
 & \innegE &\colon \NN \to \Cnl{\Env{L \dashv R}} \\
\inposO &\colon \PP \to \Cspt{\Obl{L \dashv R}}\varpi
 & \inposE &\colon \PP \to \Cpt{\Env{L \dashv R}}\mathpunct.
\end{align*}
These are given on morphisms by
\begin{align*}
\inneg\Big(n \xrightarrow{g} m\Big)^\flat &\defequiv LRn \xrightarrow{\epsilon_n \dcomp f} m
& \inpos\Big(p \xrightarrow{f} q\Big)^\flat &\defequiv Lp \xrightarrow{Lf} Lq \\
\inneg\Big(n \xrightarrow{g} m\Big)^\sharp &\defequiv Rn \xrightarrow{Rf} Rm
& \inpos\Big(p \xrightarrow{f} q\Big)^\sharp &\defequiv p \xrightarrow{f \dcomp \eta_q} RLq \mathpunct.
\end{align*}
\end{lemma}
\begin{proof}
The linearity and thunkability conditions follow from naturality of
\(\epsilon\) and \(\eta\) (by \autoref{lem:envelope-thunkable-char}).  The functor
laws follow from those of \(R\) and \(L\).
\qed
\end{proof}

\begin{lemma}[{\lemmaref{UniMath/CategoryTheory/Nonassociative/Duploids/FromAdjunction/Oblique.v}{470}{isweq\_on\_ob\_negative\_to\_oblique\_duploid}}]
\label{lem:isweq-on-objects-obl-incl}
The inclusion functors \(\innegO\) and \(\inposO\) are equivalences on
objects.
\qed
\end{lemma}

\begin{lemma}[{\lemmaref{UniMath/CategoryTheory/Nonassociative/Duploids/FromAdjunction/Envelope.v}{904}{split\_eso\_negative\_to\_envelope\_duploid}}]
\label{lem:isweq-eso-env-incl}
The inclusion functors \(\innegE\) and \(\inposE\) are split essentially
surjective.
\qed
\end{lemma}

\autoref{lem:isweq-on-objects-obl-incl} does not generalize to the envelope
duploid directly; only certain negative envelope objects are of the
form \(\innegE n\), even if \(\NN\) and \(\PP\) are univalent (by
\autoref{rem:envelope-syn-polarities-char}):

\begin{lemma}[{\lemmaref{UniMath/CategoryTheory/Nonassociative/Duploids/FromAdjunction/EqualizingRequirement.v}{756}{envelope\_chosen\_negative\_ob\_eq}}]
\label{lem:envelope-ob-eq-negative}
Let \(\NN\) and \(\PP\) be univalent. Every object \(a : \Env{L \dashv
 R}\) with a negative polarity choice is identified with
 \(\upshift{a} \equiv \inneg(\projneg{a})\). Dually, every object
 \(b : \Env{L \dashv R}\) with a positive polarity choice is
 identified with \(\downshift{b} \equiv \inpos(\projpos{a})\).
\end{lemma}
\begin{proof}
We show the former. We need to identify
\begin{equation}
\Big(\projpos a \xrightarrow{\crossmor a} \projneg a\Big)
= \Big(R(\projneg a) \xrightarrow{\crossmor(\upshift a)} \projneg a\Big)\mathpunct.
\label{eq:envelope-ob-eq-negative}
\end{equation}
By assumption, \(\crossmor a^\sharp : \morsof\PP{\projpos
a}{R(\projneg a)}\) is an isomorphism. Thus by \autoref{lem:fun-thm-of-ids} we
have \(\langle \projpos a, \crossmor a^\sharp \rangle
=_{\ifan\PP{R(\projneg a)}} \langle R(\projneg a), \id_{R(\projneg a)}
\rangle\).  Since \(\crossmor a^\sharp = \id_{R(\projneg a)}\) and
\(\id_{R(\projneg a)} \equiv \crossmor(\upshift a)^\sharp\),
\autoref{eq:envelope-ob-eq-negative} follows from injectivity of
\(\sharp\).
\qed
\end{proof}

We now state the equalizing requirement, which solves our remaining problems.

\begin{theorem}[{\lemmaref{UniMath/CategoryTheory/Nonassociative/Duploids/FromAdjunction/EqualizingRequirement.v}{579}{ff\_negative\_to\_oblique\_iff\_equ}}]
\label{thm:equalizing}
The following are equivalent:
\begin{enumerate}
\item \needspace{2\baselineskip}\label{itm:negeqfst}The functor \(\innegO \colon \NN \to \Csnl{\Obl{L \dashv R}}\varpi\) is fully\hyp{}faithful.\label{itm:negOFF}
\item \phantomsout{The functor \(\innegO \colon \NN \to \Csnl{\Obl{L \dashv R}}\varpi\)} is an isomorphism of categories.\label{itm:negOEQ}
\item \needspace{2\baselineskip}The functor \(\innegE \colon \NN \to \Cnl{\Env{L \dashv R}}\) is
fully\hyp{}faithful.\label{itm:negEFF}
\item \phantomsout{The functor \(\innegE \colon \NN \to \Cnl{\Env{L \dashv R}}\)}
is a strong equivalence of
categories.\label{itm:negeqstreqv}\label{itm:negEEQ}\footnote{Conditions \ref{itm:negeqstreqv} and \ref{itm:poseqstreqv} are
not propositions in general, so ``equivalent'' means ``if and only if''
rather than ``equivalence of types'' for them.\label{fn:equalizing-equiv-iff}}
\item \label{itm:negeqlast}\label{itm:negUP}For all \(n : \NN\), the fork
\(\begin{tikzcd}[cramped]
            LRLRn & LRn & n
            \arrow["{LR\epsilon_n}", shift left=2, from=1-1, to=1-2]
            \arrow["{\epsilon_{LRn}}"', shift right=2, from=1-1, to=1-2]
            \arrow["{\epsilon_n}", from=1-2, to=1-3]
    \end{tikzcd}\)
is a coequalizer.
That is, for all objects \(n, m : \NN\) and morphisms \(f \colon
    LRn \to m\) satisfying \(LR\epsilon_n \dcomp f = \epsilon_{LRn}
    \dcomp f\), there exists a unique morphism \(f^* \colon n \to m\) such that
\(\epsilon_n \dcomp f^* = f\).
\end{enumerate}
\end{theorem}
\begin{proof}
The implications \((\ref{itm:negOEQ}) \Rightarrow
(\ref{itm:negOFF})\) and \((\ref{itm:negEEQ}) \Rightarrow
(\ref{itm:negEFF})\) are immediate. Their converses
\((\ref{itm:negOFF}) \Rightarrow (\ref{itm:negOEQ})\) and
\((\ref{itm:negEFF}) \Rightarrow (\ref{itm:negEEQ})\) follow from
Lemmas~\ref{lem:isweq-on-objects-obl-incl} and
\ref{lem:isweq-eso-env-incl}.  The equivalences \((\ref{itm:negOFF})
\Leftrightarrow (\ref{itm:negUP})\) and \((\ref{itm:negEFF})
\Leftrightarrow (\ref{itm:negUP})\) follow from the
characterizations of linear morphisms in terms of the counit
\(\epsilon\) of the adjunction
(\autoref{lem:envelope-thunkable-char}).
\qed
\end{proof}

\begin{theorem}[{dual of \ref{thm:equalizing}}]
Dually, the following are equivalent:
\begin{enumerate}[label=\Alph*., ref=\Alph*]
\item \needspace{2\baselineskip}\label{itm:poseqfst}The functor \(\inposO \colon \PP \to \Cspt{\Obl{L \dashv R}}\varpi\) is fully\hyp{}faithful.\label{itm:posOFF}
\item \phantomsout{The functor \(\inposO \colon \PP \to \Cspt{\Obl{L\dashv R}}\varpi\)} is an isomorphism of categories.\label{itm:posOEQ}
\item \needspace{2\baselineskip}The functor \(\inposE \colon \PP \to \Cpt{\Env{L \dashv R}}\) is fully\hyp{}faithful.\label{itm:posEFF}
\item \phantomsout{The functor \(\inposE \colon \PP \to \Cpt{\Env{L \dashv R}}\)}
is a strong equivalence of
categories.\label{itm:posEEQ}\label{itm:poseqstreqv}\footref{fn:equalizing-equiv-iff}
\item \label{itm:poseqlast}\label{itm:posUP}For all \(p : \PP\), the fork
\(\begin{tikzcd}[cramped] 
  	        q & RLq & RLRLq
  	        \arrow["{\eta_p}"', from=1-1, to=1-2]
  	        \arrow["{RL\eta_q}", shift left=2, from=1-2, to=1-3]
  	        \arrow["{\eta_{RLq}}"', shift right=2, from=1-2, to=1-3]
  \end{tikzcd}\)
is an equalizer.
That is, for all objects \(p, q : \PP\) and morphisms \(g \colon p \to RLq\)
satisfying \(g \dcomp RL\eta_q = g \dcomp \eta_{RLq}\), there
exists a unique morphism \(g^* \colon p \to q\) such that \(g^* \dcomp
  \eta_q = g\).
\qed
\end{enumerate}
\end{theorem}

\begin{definition}[{\lemmaref{UniMath/CategoryTheory/Nonassociative/Duploids/FromAdjunction/EqualizingRequirement.v}{86}{is\_negative\_equalizing}}]
\label{def:equalizing}
An adjunction satisfying conditions
\ref{itm:negeqfst}--\ref{itm:negeqlast} of \autoref{thm:equalizing} is
\emph{negative equalizing}, and an adjunction satisfying
conditions \ref{itm:poseqfst}--\ref{itm:poseqlast} is \emph{positive
equalizing}. An adjunction which satisfies conditions
\ref{itm:negeqfst}--\ref{itm:negeqlast} \textbf{and} conditions
\ref{itm:poseqfst}--\ref{itm:poseqlast} is \emph{fully equalizing}.
\end{definition}

\begin{exampledef}[{\cite[\sect{} 4.2]{MunchMaccagnoni14}}]
A duploid's \(\upshift \dashv \downshift\) adjunction (\autoref{rem:dupl-adjunction}) is fully equalizing.
\end{exampledef}

With the equalizing requirement, we can prove univalence for both the
oblique and envelope duploids. The former is simple, the latter needs
the converse of \autoref{lem:envelope-syn-valid}.

\begin{theorem}[{\lemmaref{UniMath/CategoryTheory/Nonassociative/Duploids/FromAdjunction/EqualizingRequirement.v}{740}{is\_univalent\_oblique\_duploid\_iff}}]
\label{thm:oblique-univalent}
If \(L \dashv R\) is fully equalizing, then the oblique duploid
\(\Obl{L \dashv R}\) is univalent as a split duploid if and only if
\(\NN\) and \(\PP\) are univalent.
\end{theorem}
\begin{proof}
Recall that univalence as a split duploid means that \(\Csnl{\Obl{L
\dashv R}}\varpi\) and \(\Cspt{\Obl{L \dashv R}}\varpi\) are
univalent. Using \autoref{lem:univalence-over-catiso}, univalence may be
transported (in both directions) across the isomorphisms \(\innegO :
\NN \cong \Csnl{\Obl{L \dashv R}}\varpi\) and \(\inposO : \PP \cong
\Cspt{\Obl{L \dashv R}}\varpi\).
\qed
\end{proof}

\begin{lemma}[{\lemmaref{UniMath/CategoryTheory/Nonassociative/Duploids/FromAdjunction/EqualizingRequirement.v}{252}{is\_negative\_fixed\_point\_from\_pre\_fixed\_point}}]
\label{lem:pre-weq-fixed-point}
If \(L \dashv R\) is negative equalizing, then for all \(n : \NN\),
the equation \(LR\epsilon_n = \epsilon_{LRn}\) implies that
\(\epsilon_n\) is an isomorphism. Dually, if \(L \dashv R\) is
positive equalizing, then for all \(p : \NN\), the equation \(RL\eta_p
= \eta_{RLp}\) implies that \(\eta_p\) is an isomorphism.
\qed
\end{lemma}

\begin{lemma}[{\lemmaref{UniMath/CategoryTheory/Nonassociative/Duploids/FromAdjunction/EqualizingRequirement.v}{794}{envelope\_chosen\_negative\_from\_negative}}]
\label{lem:chosen-negative-iff-negative}
If \(L \dashv R\) is positive equalizing, then every semantically
negative object in \(\Env{L \dashv R}\) has a negative polarity
choice. Dually, if \(L \dashv R\) is negative equalizing, then every
semantically positive object has a positive polarity choice.
\end{lemma}
\begin{proof}
We show the negative case. Let \(a : \Env{L \dashv R}\) be a negative
object with a positive polarity choice, so that \(\crossmor
a^\flat\) is an isomorphism.
Then \(\crossmor a^\sharp =
\varphi_{(\projpos a),(\projneg a)}(\crossmor a^\flat) \equiv
\eta_{\projpos a} \dcomp R(\crossmor a^\flat)\) is the composite of
two isomorphisms: \(R(\crossmor a^\flat)\) (by assumption) and
\(\eta_{\projpos a}\) (by \autoref{lem:envelope-polarities-char} and
\autoref{lem:pre-weq-fixed-point}).
\qed
\end{proof}

From Lemmas~\ref{lem:envelope-ob-eq-negative} and
\ref{lem:chosen-negative-iff-negative} we obtain that \(\innegE\) and
\(\inposE\) are isomorphisms. Univalence of the envelope duploid then
follows from Theorems~\ref{lem:univalence-over-catiso} and
\ref{thm:duploid-univ-of-subcats}:

\begin{theorem}[{\lemmaref{UniMath/CategoryTheory/Nonassociative/Duploids/FromAdjunction/EqualizingRequirement.v}{873}{is\_catiso\_negative\_to\_envelope\_duploid}}]
\label{lem:envelope-inj-isos}
If \(L \dashv R\) is fully equalizing, and \(\NN\) and \(\PP\) are
univalent, then \(\innegE \colon \NN \to \Cnl{\Env{L \dashv R}}\) and
\(\inposE \colon \PP \to \Cpt{\Env{L \dashv R}}\) are isomorphisms of
categories.
\qed
\end{theorem}

\begin{theorem}[{\lemmaref{UniMath/CategoryTheory/Nonassociative/Duploids/FromAdjunction/EqualizingRequirement.v}{914}{is\_univalent\_envelope\_duploid}}]
\label{thm:envelope-univalent}
If \(L \dashv R\) is fully equalizing, and \(\NN\) and \(\PP\) are
univalent, then the envelope duploid \(\Env{L \dashv R}\) is
univalent.
\qed
\end{theorem}
\section{Conclusions and Future Work}
\label{sec:conclusions}
The contributions of this work are twofold. First, we have formalized
part of Munch\hyp{}Maccagnoni's work on duploids \cite{MunchMaccagnoni14,MunchMaccagnoni13} in \textsc{Rocq}. The formalization covers
unital magmoids and two variants of duploids: single\hyp{}sorted duploids
and split duploids. Second, we have made novel contributions to their
study in univalent foundations. We have defined what it means to be
univalent for unital magmoids, and for both variants of duploids. We
studied two versions of the duploid arising from an adjunction: the
original ``oblique duploid'' construction of Munch\hyp{}Maccagnoni
\cite{MunchMaccagnoni14} which is naturally a split duploid, and a
novel ``envelope duploid'' construction which is a single\hyp{}sorted
duploid. Finally, we have shown both of these to be univalent -- in
their respective senses -- when arising from a fully equalizing
adjunction between univalent categories.

Our formalization of Munch\hyp{}Maccagnoni's work ends at
showing that each duploid is equivalent to one arising from an
adjunction. This result is the first part of the full ``structure
theorem'' \cite{MunchMaccagnoni14}: that the category of duploids is
equivalent to the category of fully equalizing adjunctions. We note
that the hom\hyp{}set law of categories prevents even defining the
1-category of duploids in univalent foundations, and suggests that
duploids ought to be treated as a 2-category instead. Future work
could include a study of the 2-category of duploids, and a
2-categorical structure theorem for duploids.

We fall short of defining a Rezk completion for duploids: a weakly
equivalent univalent duploid for any duploid \(\D\). One way could be
to use \autoref{thm:envelope-structure}, but Rezk\hyp{}complete the categories
\(\Cnl\D\) and \(\Cpt\D\) before assembling \(\Env{\upshift \dashv
\downshift}\). This procedure eluded our formalization due to
complexity arising from the 2-categorical nature of
adjunctions.

Lastly, the consequences of univalence of unital magmoids could be
further studied. For example, while we proved directly that polarity
shifts are unique in a univalent duploid, future work could study
other universal properties in univalent unital magmoids.
\bibliographystyle{splncs04}
\bibliography{bibliography}
\label{lastpage}
\end{document}